\documentclass[fleqn,12pt,twoside]{article}
\usepackage{espcrc1}
\usepackage{epsfig}

\title{Model for dilepton rates from a fireball\thanks{Work supported in part by BMBF and GSI.}}

\author{R.A. Schneider\address[TUM]{Physik Department, Technische Universit\"{a}t M\"{u}nchen \\
D-85747 Garching, Germany}, T. Renk\addressmark and W. Weise\addressmark}

\begin{document}

\maketitle

\begin{abstract}
We calculate the dilepton emission rate from a fireball created in an ultra-relativistic heavy-ion
collision. For the partonic phase, we complement the perturbative results by a phenomenological
approach based on lattice QCD results, whereas in the hadronic phase temperature and finite baryon
density effects on the spectral function are considered. The resulting rates are compared to data from
CERES/NA45.
\end{abstract}
\section{Introduction}
From lattice simulations it is known that the theory of strong interactions, QCD, undergoes a phase
transition at a temperature $T_C \sim$ 160 MeV \cite{FK00} from a confined hadronic
phase to a phase where quarks and gluons constitute the relevant degrees of freedom, the quark-gluon
plasma (QGP).
It is hoped that it is possible to create this QGP in ultrarelativistic heavy-ion collisions at CERN
and RHIC. Dileptons ($e^+e^-$ and $\mu^+ \mu^-$ pairs) are interesting probes in this context since
they do not interact strongly but rather escape unthermalized from the hot and dense region formed in
such collisions, the fireball. In the early stage, a thermalized QGP phase is expected to exist. There,
dileptons originate mainly from $q\bar{q}$ annihilation processes. As the fireball expands, it cools
off and hadronization sets in at $T_C$. In the hadronic phase, the main dilepton sources are pion and
kaon annihilation processes which are enhanced due to the formation of the light vector mesons $\rho,
\omega$ and $\phi$.\\
The differential dilepton emission rate from a hot domain in thermal equilibrium is given by
\begin{equation}
\label{E-Rate} \frac{dN}{d^4xd^4q} = \frac{\alpha^2}{\pi^3q^2}\frac{1}{e^{q^0/T} -1} \mbox{Im}
\bar{\Pi}(q,T)
\end{equation}
where $\bar{\Pi}$ denotes the spin-averaged time-like self energy of a photon in the heat bath, $q$ is
the photon four-momentum, $\alpha = \frac{e^2}{4\pi}$, and the lepton masses have been neglected. This
rate has to be integrated over the time evolution of the fireball volume $V(t)$ and temperature $T(t)$.
\section{The QGP phase}
Thermal perturbation theory is presumably not applicable to evaluate Im$\bar{\Pi}$ in the QGP. The
coupling constant $g_s \sim 2$ is large, and the maximal temperatures reached in ultrarelativistic heavy ion collisions  are still
$\mathcal{O}(T_C)$, indicating that non-perturbative physics is important. Consequently, we try to use
a more phenomenological approach to thermal QCD. It has been shown that it is possible to describe the
equation of state (EoS) of QCD as obtained from lattice calculations by the EoS of a free gas of
quasiparticles with thermally generated masses $m(T)$ \cite{PKS}. The model employs 'effective', non-interacting massive
quarks and gluons as the relevant degrees of freedom, all dynamics is
incorporated in a $T$-dependent effective coupling constant $g_s(T)$. This coupling is based on the expression for
 the running coupling at $T=0$. For the quark masses $m_q(T)$ the two parameters $T_s$ and $\lambda$ are fitted to the
lattice data:
\begin{equation}
m_q^2(T) =  \frac{1}{3} g_s^2(T) T^2 \quad \mbox{with} \quad g_s^2(T) = \frac{48 \pi^2}{(11 N_C - 2
N_F) \ln\left[ (T + T_s) / (T_C/\lambda) \right]^2}.\label{m_q}
\end{equation}
Hence, the only QGP contribution to Im$\bar{\Pi}$ in eq.(\ref{E-Rate}) is the $q\bar{q}$-loop,
evaluated with the thermal quark masses (\ref{m_q}).
\section{The hadronic phase}
\begin{figure}[h!]
\epsfig{file=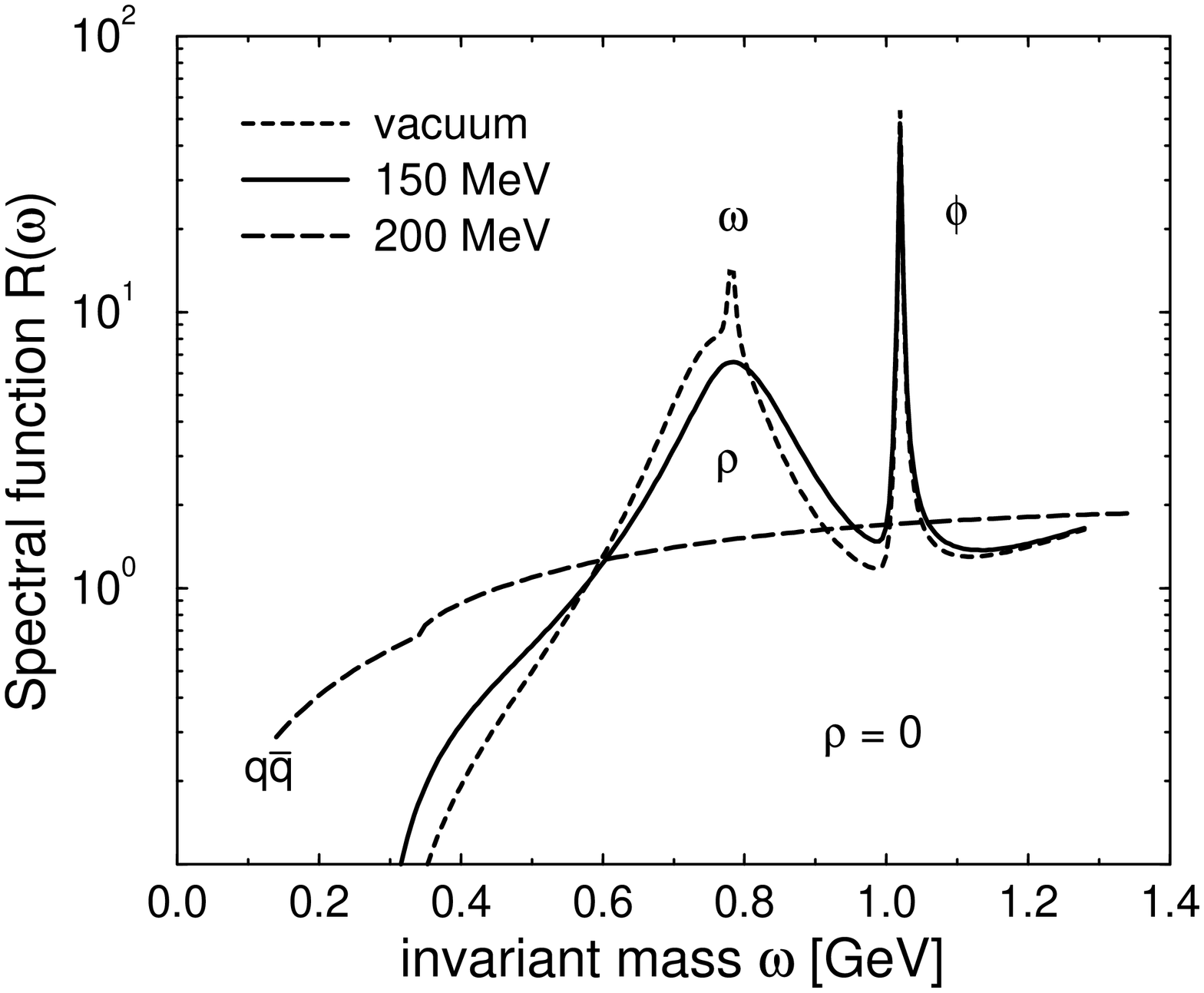,width=6.5cm, angle = -90} \epsfig{file=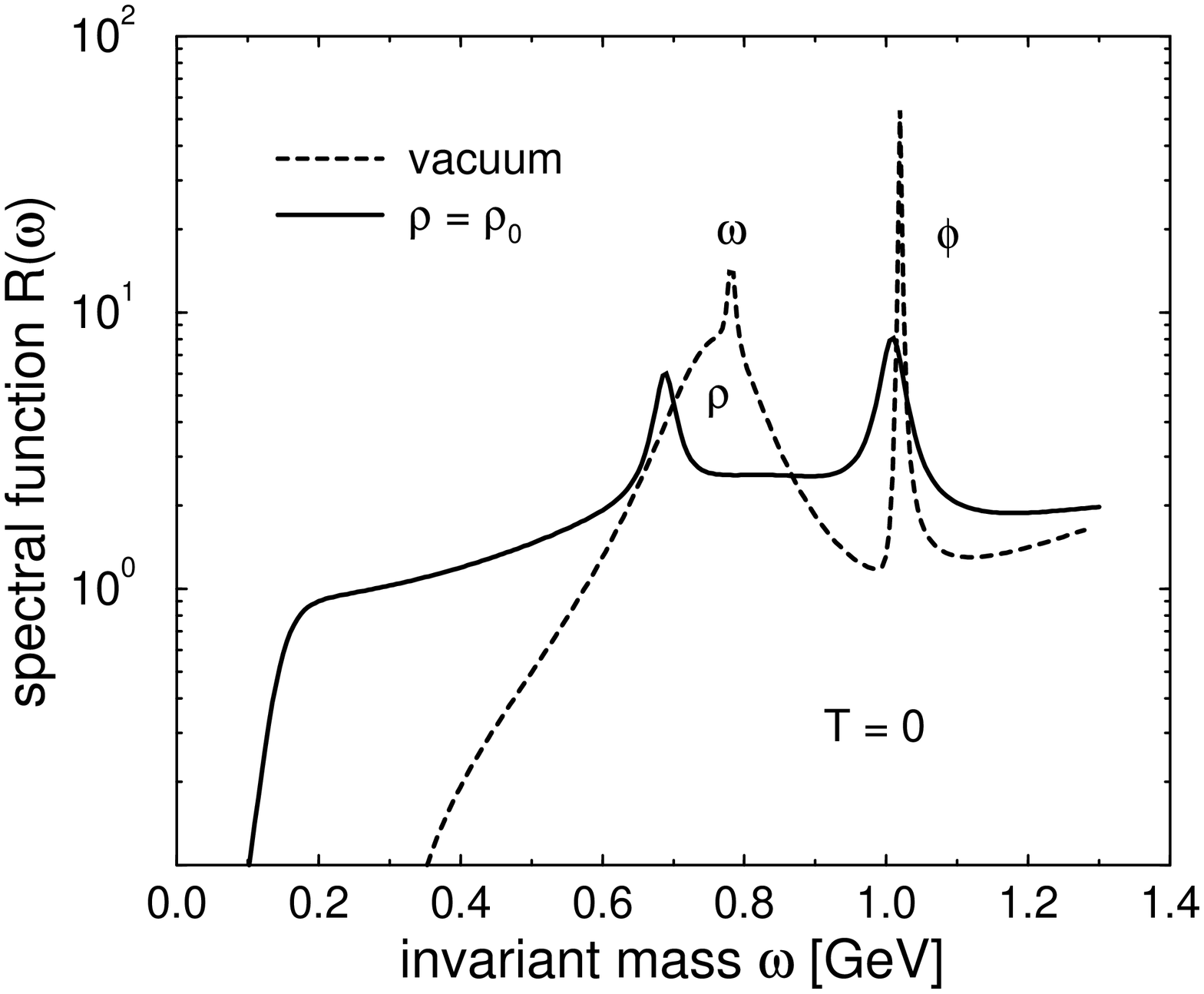,width=6.5cm,
angle = -90} \caption{The photon spectral function $R(\omega) = -12\pi/\omega^2 \ \mbox{Im}
\bar{\Pi}(\omega)$ at finite temperature and $\rho = 0$ (left panel) and at $T=0$ and finite density
(right panel). The $q\bar{q}$ line in the left panel shows the spectral function in the QGP phase for
massless quarks.} \label{spectra}
\end{figure}

Below $T_C$, confinement sets in and the effective degrees of freedom change. The photon couples now to
hadronic $J^P = 1^-$ states: the $\rho$, $\omega$ and $\phi$ mesons and multi-pion states carrying the
same quantum numbers. The electromagnetic current-current correlation function can be connected to the
currents generated by these mesons using an effective Lagrangian which approximates the $SU(3)$ flavour
sector of QCD at low energies. The appropriate model for our purposes is the {\em improved Vector Meson
Dominance} model combined with chiral dynamics of pions and kaons as described in \cite{KKW1}. Within
this model, the following relation between the imaginary part of the irreducible photon self-energy
$\mbox{Im} \bar{\Pi}$ and the vector meson self-energies $\Pi_V(q)$ in vacuum is derived:
\begin{equation}
\mbox{Im} \bar{\Pi}(q) = \sum \limits_V \frac{\mbox{Im} \Pi_V(q)}{g_V^2} \ |F_V(q)|^2, \label{ImBarPi}
\quad F_V(q) = \frac{\left( 1- \frac{g}{g^0_{V}} \right)q^2 - m_V^2}{q^2 - m_V^2 + i
\mbox{Im}\Pi_V(q)},
\end{equation}
where $m_V$ are the (renormalized) vector meson masses, $g^0_V$ is the $\gamma V$ coupling and $g$ the
$\Phi V$ coupling, where $\Phi$ stands for one of the pseudoscalar Goldstone bosons $\pi^\pm, \pi^0$
and $K^\pm, K^0$. Eq.(\ref{ImBarPi}) is valid for a virtual photon with vanishing three-momentum.
Finite temperature modifications of the vector meson self-energies are calculated using thermal Feynman
rules, the explicit calculations for the $\rho$-, $\omega$- and $\phi$-meson can be found in
\cite{SW00}. At SPS energies, there is still considerable stopping of the interpenetrating nuclei,
resulting in a net baryon density in the central rapidity region. For the evaluations of these finite
baryon density effects, we use the results discussed in \cite{KKW2}. There it was shown that in the
linear density approximation $\Pi_V$ is related to the vector meson - nucleon scattering amplitude
$T_{VN}$
\begin{equation}
\Pi_V(q^0, \vec{q} = 0; \rho) = \Pi_V^{vac} - \rho T_{VN}(q),\quad T_{VN}(q) = - \frac{i}{3} \int d^4 x
 \ e^{iqx} \langle N | \mathcal{T} j_\mu(x) j^\mu(0) | N \rangle, \label{PiV}
\end{equation}
with $|N\rangle$ being the nucleon states with vanishing momentum. In the following, we assume that the
temperature- and density dependences of $\Pi_V$ factorize, {\em i.e.} we replace in eq.(\ref{PiV})
$\Pi_V^{vac}$ by the temperature-dependent $\Pi_V(T)$ and leave $T_{VN}$ unaffected. This amounts to
neglecting contributions from matrix elements like $\langle N\pi | \mathcal{T} j_\mu(x) j^\mu(0) | \pi
N \rangle$ (meson-nucleon-pion scatterings where the pion comes from the heat bath). Furthermore, this
approximation does not take into consideration a possible $T$-dependent pion or nucleon mass. The
corresponding spectra at finite temperature and density are depicted in figure \ref{spectra}.
\section{Dilepton rates}
\begin{figure}[t]
\begin{center}
\epsfig{file=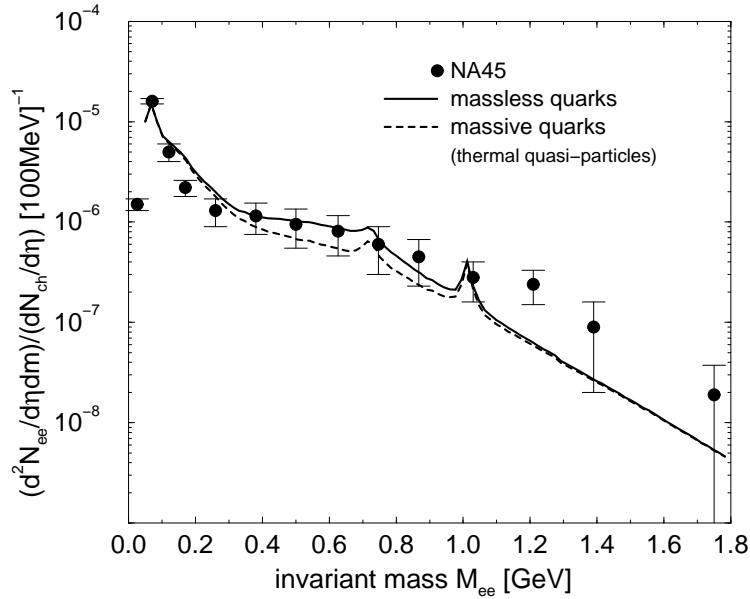,width=8cm, angle = -90} \label{CERES1} \caption{Dilepton rates 
as a function of the invariant $e^+e^-$ mass $M$, calculated in a mixed scenario
 (hadrons below $T_C$, quarks above $T_C$ = 150 MeV) for 158 AGeV Pb-Au collisions at CERES.}
\end{center}
\end{figure}

Eq.(\ref{E-Rate}) is now integrated over the space-time evolution of the fireball and the detector
acceptance of the CERES/NA45 experiment at CERN, using the expressions for Im$\bar{\Pi}$ from the
previous sections. We use simple parametrizations, adapted to the CERES conditions, for the volume
$V(t)$ and the temperature $T(t)$ \cite{SW00} based on more involved calculations utilizing microscopic
transport equations. The corresponding rate and the data \cite{CERES} are shown in figure \ref{CERES1}.
We have considered two cases for the QGP phase: a scenario with massive quarks according to eq.(\ref{m_q})
and an ideal gas scenario with massless quarks. \\
The low invariant mass region 0.3 - 0.6 GeV is mainly filled by the strong broadening of the
$\rho$-meson at finite density, contributions from the early QGP phase are subdominant. In the quasiparticle approach, the quark masses appear to become heavy close to $T_C$ ($m_q \sim 0.3 - 0.5 $ GeV), thus there is no radiation from the QGP below the $2m_q$ threshold in the invariant mass.\\
The $\omega$ meson almost loses its quasiparticle structure at high temperatures $T \geq 120$ MeV because of the
scattering $\omega \pi \rightarrow \pi\pi$ off thermally excited pions. Its peak structure is not
visible in the dilepton yield and this may act as a signal for medium effects when the corresponding
experimental resolution is increased. On the contrast, the $\phi$ meson remains a distinct peak over
the flat QGP spectrum and can be used as both a thermometer and a chronometer for the fireball
expansion. In the high invariant mass region, the contribution from the hadronic phase dies away, and
the slope of the dilepton yield is mainly dominated by the averaged temperature $ - \langle 1/T
\rangle$ and is insensitive to the existence of thermal masses. We also checked that a possible mixed phase would not leave any distinct traces in the dilepton spectrum. \\
To uniquely identify signals for the formation of a QGP, it remains a crucial task to further investigate temperature and density effects on the hadronic spectra.

\end{document}